\documentclass{jfm}
\usepackage{graphicx}
\usepackage{amsmath,amssymb}
\usepackage{epstopdf, epsfig}

\shorttitle{Effects of slippage on the dewetting of a droplet}
\shortauthor{T. S. Chan, J. D. McGraw, T. Salez,  R. Seemann, M. Brinkmann}

\title{Effects of slippage on the dewetting of a droplet}

\author{Tak Shing Chan\aff{1}
  \corresp{\email{tak.chan@physik.uni-saarland.de}},
  Joshua D. McGraw\aff{2},
  Thomas Salez\aff{3,4},
  Ralf Seemann\aff{1}
 \and Martin Brinkmann\aff{1}}

\affiliation{\aff{1}Experimental Physics, Saarland University, D-66041 Saarbr\"ucken, Germany
\aff{2}D\'epartement de Physique, Ecole Normale Sup\'erieure/PSL Research University, CNRS, 24 rue Lhomond, 75005 Paris, France\\
\aff{3}Laboratoire de Physico-Chimie Th\'eorique, UMR CNRS Gulliver 7083, ESPCI Paris, PSL Research University, Paris, France
\aff{4}Global Station for Soft Matter, Global Institution for Collaborative Research and Education, Hokkaido University, Sapporo, Japan}

\begin{document}
\maketitle

\begin{abstract}
\label{gen:abstr}
In many macroscopic dynamic wetting problems, it is assumed that the macroscopic interface is quasistatic, and the dissipation appears only in the region close to the contact line. When approaching the moving contact line, a microscopic mechanism is required to regularize the singularity of viscous dissipation. On the other hand, if the characteristic size of a fluidic system is reduced to a range comparable to the microscopic regularization length scale, the assumption that viscous effects are localized near the contact line is no longer justified. In the present work, such microscopic length is the slip length. Our recent study on dewetting polymer microdroplets demonstrated that slip plays a dominant role in the shape evolution as the droplet relaxes toward equilibrium \citep{McGraw16}. The transient profiles of the droplet were found to be highly non-spherical, meaning that the evolution is not quasistatic. In the present theoretical study, we investigate the dewetting of a droplet using the boundary element method. Specifically, we solve for the axisymmetric Stokes flow with i) the Navier-slip boundary condition at the solid/liquid boundary, and ii) a time-independent microscopic contact angle at the contact line. The profile evolution is computed for different slip lengths and equilibrium contact angles. When  decreasing the slip length, the typical nonsphericity first increases, reaches a maximum at a characteristic slip length $\tilde{b}_m$, and then decreases. Regarding different equilibrium contact angles, two universal rescalings are proposed to describe the behavior for slip lengths larger or smaller than $\tilde{b}_m$. Around $\tilde{b}_m$, the early time evolution of the profiles at the rim can be described by similarity solutions. The results are explained in terms of the structure of the flow field governed by different dissipation channels: viscous elongational flows for large slip lengths, friction at the substrate for intermediate slip lengths, and viscous shear flows for small slip lengths. Following the transitions between these dominant dissipation mechanisms, our study indicates a crossover to the quasistatic regime when the slip length is small compared to the droplet size.
\end{abstract}

\section{Introduction}\label{intro}

A classical problem of dynamic wetting is the spreading of a droplet when it is placed in contact with a smooth and chemically homogeneous substrate \citep{C88,BEIMR09}. For complete wetting, with a vanishing equilibrium contact angle, the spreading process follows the well-known Tanner's law \citep{V76,T79} stating that the contact line radius $R$ grows in time $t$ as a power law $R\sim t^{1/10}$. This asymptotically valid relation is derived with the assumption that the droplet maintains a spherical-cap shaped profile during spreading, except in the vicinity of the moving contact line, where the interface is deformed strongly due to viscous effects. The general assumptions of a quasistatic macroscopic interface profile and a steady viscous flow in the region close to the contact line have been central guidelines in studies of dynamic wetting problems \citep{BEIMR09,Snoeijer13,Sui2014}. Examples include industrial applications such as oil recovery \citep{Sahimi93}, immersion lithography \citep{WPER11} and coating \citep{WeRu04}, as well as natural phenomena \citep{BEIMR09} such as liquid droplets sliding on the surface of a leaf. The basis of these assumptions lies in the wide separation of length scales between the extension of the interface and a microscopic length. As specifically discussed here, this microscopic length scale may be the slip length. In the cases where a no-slip condition is assumed for the solid/liquid boundary, other microscopic length scales in specific models have been proposed to relax the singularity of infinite viscous dissipation \citep{HS71} at the contact line as reviewed in \citet{BEIMR09,Snoeijer13,Sui2014}. 

There have been extensive studies on the measurement of the slip length due to the development of new experimental techniques \citep{neto05RPP,bocquetSCR09,Guo2013}. Interestingly, in some studies using polymer melts as working fluids, slip lengths as large as a few micrometers have been reported  \citep{Reiter2001,leger03JPCM,FJMWW05,fetzer07LMR_thr,baeumchen09PRL,Haefner2015}. These findings raise fundamental questions on the description of the contact line motion and the evolution of the interface profile, particularly in micrometric \citep{cuenca13PRL,Setu2015} or nanometric \citep{Falk2010} systems, for which the separation of length scales may not be fulfilled. 

A recent experimental and theoretical study on dewetting polymer microdroplets \citep{McGraw16} showed that the transient droplet shape evolution, in the regime where the slip length is comparable to or larger than the typical droplet size, is much richer than one expects under the assumptions of quasistatic profiles and dissipation localized near the contact line. The transient droplet profiles are indeed found to be non-spherical (i.e. non-quasistatic), and highly dependent on the precise value of the slip length. One characteristic feature of the dewetting process is the development of a transient ridge for relatively small slip lengths, which are nevertheless comparable to the droplet size. The ridge was found to be more pronounced when the slip length is smaller and avoided for larger slip lengths due to elongational flow inside the droplet. On the other hand, as discussed above, when the slip length is many orders of magnitude smaller than the droplet size, one expects to recover the typical quasistatic sequence of spherical cap shaped profiles \citep{BEIMR09}.

In this article, by extending the theoretical work of \citet{McGraw16}, we elucidate the transition between the quasistatic and non-quasistatic evolutions of a dewetting droplet. We study the dewetting of a viscous droplet for a wide range of slip lengths and various equilibrium contact angles using the boundary element method. The non-sphericity of the droplet increases when the slip length is first decreased from the full slip limit. Further decreasing the slip length, we observe a new feature with respect to previous works \citep{McGraw16}: the non-sphericity reaches a maximum and then starts to decrease. This behavior is demonstrated for different equilibrium contact angles. We give explanations for these results in terms of flow structures and the spreading of a localized ridge.

\section{Formulation}\label{form}
As an initial condition, we consider a spherical cap shaped droplet sitting on a plane and smooth substrate with a contact angle $\theta_i$, which is smaller than the equilibrium contact angle $\theta_e$. In order to minimize the surface energy, the droplet starts to retract and approaches  a spherical cap with the equilibrium contact angle. Because of the homogeneous and planar substrate, the shape of the droplet remains axisymmetric during its evolution. The droplet profile is described by the height, $h(r,t)$, of the liquid with respect to the substrate as a function of the radial distance from the central axis $r$ and time $t$. We further assume the liquid inside the droplet to be a highly viscous and incompressible Newtonian liquid so that the flow obeys Stokes equation, for which viscous effects dominate over inertial effects. The Stokes equation is given as 
\begin{eqnarray}\label{stokes1}
\eta \nabla^2 \boldsymbol{u}-\boldsymbol{\nabla} p=0\, ,
\end{eqnarray}
and the continuity equation reads
\begin{eqnarray}\label{stokes2}
\boldsymbol{\nabla} \boldsymbol{\cdot}\boldsymbol{u}=0\, ,
\end{eqnarray}
where $\boldsymbol{u}$ and $p$ are the velocity field and the pressure field in the liquid respectively, and $\eta$ is the dynamic viscosity of the liquid. 

To solve for the flow fields and the evolution of the interface profile, one needs to
specify appropriate boundary conditions. First, the stress tensor $\boldsymbol{\sigma}$ in Cartesian coordinates is given as 
\begin{equation}\label{def_stress}
\sigma_{ij}=-p\delta_{ij}+\eta\left (\frac{\partial u_i}{\partial
x_j}+\frac{\partial u_j}{\partial x_i}\right)\, ,
\end{equation}
and the stress $\boldsymbol{f}$ at the boundary reads 
\begin{equation}\label{def_stress}
\boldsymbol{f}=\boldsymbol{\sigma}\boldsymbol{\cdot}{\boldsymbol{\hat{n}}}\, .
\end{equation}
Here $\boldsymbol{\hat{n}}$ is the unit vector normal to the boundary of the droplet pointing into the enclosed fluid. 

Assuming the surrounding air flow is negligible, the tangential stress vanishes at the liquid/air boundary. The normal stress $f^{{\rm free}}_\textrm{n}\equiv \boldsymbol{f}^{{\rm free}}\boldsymbol{\cdot} {\boldsymbol{\hat{n}}} $ at the free surface is balanced by the surface tension, leading to the Young-Laplace law: 
\begin{equation}\label{ka}
f^{{\rm free}}_\textrm{n}=\gamma \kappa\, ,
\end{equation}
where $\gamma$ denotes the interfacial tension and $\kappa$ the  curvature of the free surface, which is defined as 
\begin{equation}\label{ka}
\kappa=\frac{\frac{\partial^2h}{\partial r^2}}{(1+(\frac{\partial h}{\partial r})^{2})^{3/2}}+\frac{\frac{\partial h}{\partial r}}{r(1+(\frac{\partial h}{\partial r})^{2})^{1/2}}\, .
\end{equation}
Note that disjoining pressures are not considered in this model. The evolution of the interface profile is given by the kinematic condition along the free interface, that is
\begin{equation}\label{kin}
\frac{\partial h}{\partial t}=u_z-\frac{\partial h}{\partial r}u_r\, .
\end{equation}

At the solid/liquid boundary, the velocity normal to the wall
vanishes as no penetration of fluid through the solid is
allowed. Regarding the velocity component parallel to the wall $u^{{\rm wall}}_\textrm{t}\hat{\boldsymbol{r}}$, we impose a Navier-slip condition which reads
\begin{equation}\label{}
u^{{\rm wall}}_\textrm{t}=\frac{b}{\eta} f^{{\rm wall}}_\textrm{t}\, ,
\end{equation}
where $\hat{\boldsymbol{r}}$ is the unit vector in the radial direction, $ f^{{\rm wall}}_\textrm{t}\equiv\boldsymbol{f}^{{\rm wall}}\boldsymbol{\cdot} \boldsymbol{\hat{r}}$ is the shear stress at the wall and the slip length, $b$, is assumed to be a constant. To complete the hydrodynamic problem, we impose the condition that the free surface touches the wall with a finite contact angle. This angle is assumed to be the same as the equilibrium contact angle $\theta_e$, independent of the contact line velocity. Moreover, since the substrate surface is smooth and chemically homogeneous, $\theta_e$ is also independent of the contact line position. 

\subsection{Boundary element method}\label{sec_bem}
The governing equations (\ref{stokes1}) and (\ref{stokes2}) can be formulated in the form of the boundary integral equations; a method which has been used extensively to study many interfacial flow problems \citep{Pozbook}. In this approach the velocity $\boldsymbol{u}(\boldsymbol{s}_0)$ at any point $\boldsymbol{s}_0$ can be written in terms of integrals involving the stress $\boldsymbol{f}$ and the velocity on the boundary. For the axisymmetric Stokes flow problem we study in this article, the boundary integral equations \citep{Pozbook} read 

\begin{equation}\label{bem2}
u_{\alpha}(\boldsymbol{s}_0)=-\frac{A}{4\pi\eta}\int_c\bar{G}_{\alpha\beta}(\boldsymbol{s}_0,\boldsymbol{s})f_{\beta}(\boldsymbol{s})dl(\boldsymbol{s})+ \frac{A}{4\pi}\int_c\bar{T}_{\alpha\beta\zeta}(\boldsymbol{s}_0,\boldsymbol{s})u_{\beta}(\boldsymbol{s})n_{\zeta}(\boldsymbol{s})dl(\boldsymbol{s}),
\end{equation}
where the subscripts $\alpha$, $\beta$ and $\zeta$ represent either the radial ($r$) or the vertical ($z$) components in cylindrical coordinates, and $c$ is the contour line (boundary) over which the integration takes place. For the expression of the tensor components $\bar{G}_{\alpha\beta}$ and $\bar{T}_{\alpha\beta\zeta}$, we refer to the Appendix. The value of $A$ depends on the position $\boldsymbol{s}_0$. 
\begin{eqnarray}\label{eq:dmax}
A=
\left\{
\begin{array}{l l}
1/2 & \quad \textrm{for $\boldsymbol{s}_0$ inside the system enclosed by the boundary,} \\
1 & \quad \textrm{for $\boldsymbol{s}_0$ on the closed boundary} .
\end{array}
\right.
\end{eqnarray}
We note that $\bar{G}_{\alpha\beta}$ and $\bar{T}_{\alpha\beta\zeta}$ are singular at $\boldsymbol{s}=\boldsymbol{s}_0$; the integral over the singular point is thus computed analytically by expanding the tensor components in series about $\boldsymbol{s}=\boldsymbol{s}_0$ \citep{Lee1982,vanLangerich12JFM}. 

The main advantage of the boundary element method is that the velocity field is explicitly written in terms of the velocities and the stresses on the boundary. No discretization of elements inside the droplet is required to solve for the flow fields. As given from the boundary conditions, not all the velocities and stresses at the boundary are known. For example, the velocities at the free interface are unknowns. Yet, the unknown quantities can be found by solving (\ref{bem2}) for $\boldsymbol{s}_0$ on the boundary. For a numerical treatment of the problem, the contour is discretized into small elements. A system of linear equations is then obtained from (\ref{bem2}), and the unknown quantities can be computed. Once the velocities at the free surface have been computed, one can determine the profile evolution using the kinematic condition (\ref{kin}).

Initially, the droplet has a spherical cap shape with an contact angle $\theta_i$. Due to the small molecular relaxation time scale at the contact line, the contact angle quickly reaches the equilibrium contact angle $\theta_e$ microscopically \citep{BEIMR09}. To approximate this initial microscopic contact angle in our numerical computations, we assume that at $t=0$, there is a kink in the interface profile at the contact line position. The line connecting the first numerical marker point and the contact line makes an angle $\theta_e$ with the substrate. Due to this kink, the magnitude of the approximated interfacial curvature near the contact line is larger than that on the rest of the interface, thus the Laplace pressure is unbalanced and the pressure gradient initiates a flow. Hence, the contact line starts to move towards the center. 

\begin{figure}
\begin{center}
\includegraphics[width=0.8\textwidth]{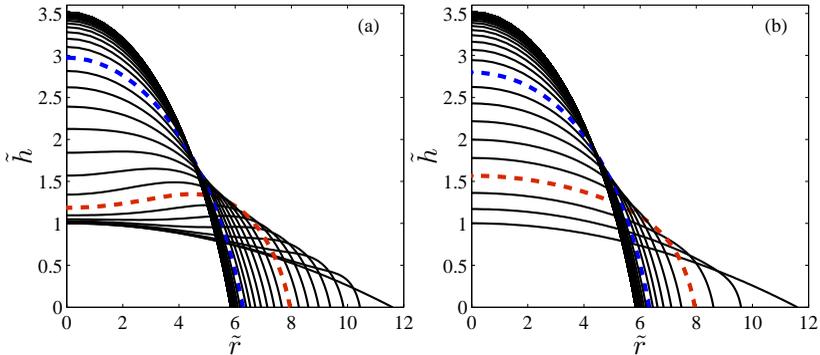}
\caption{Evolution of the droplet profiles. The initial shape of the droplet is a flat spherical cap with contact angle $\theta_i = 10\,^\circ$. The final equilibrium droplet has contact angle $\theta_e = 62\,^\circ$. Two of the profiles in each figure are plotted in dashed lines for ease of reading. (a) $\tilde{b}$ = 0.46.  The time interval $\delta \tilde{t}$ between two neighbouring curves is 3.48. (b) $\tilde{b}$ = 23.2 and $\delta \tilde{t}$ = 1.28. }\label{profiles}
\end{center}
\end{figure}

We nondimensionalize the problem as follows: all lengths are rescaled by the initial maximum height of the droplet $h_0$, and all the times by the viscous capillary time scale $h_0\eta/\gamma$. All these dimensionless variables are denoted with a tilde.  We are thus left with three independent dimensionless parameters. In the following, we consider the initial contact angle $\theta_i$, the equilibrium contact angle $\theta_e$ and the rescaled slip length $\tilde{b}\equiv b/h_0$ as the control parameters. For all our numerical computations, 300 marker points are used to describe the interface profile of the droplet. The vertical separation between two marker points is approximately 0.003. For smaller separations, the profile evolution becomes unstable. We then set the smallest rescaled slip length to $\tilde{b} = 0.023$, which is about ten times the marker separation. Hence for all our computations, the rescaled slip length is varied in a range $\tilde{b}>0.023$.

\section{Results and discussion}\label{result}
In this Section, we present the results of our numerical computations. In section \ref{profilevol}, we revisit the interfacial profile evolution as studied by \citet{McGraw16}. We characterize and quantify the deviation of the transient droplet profiles from a spherical cap. Then we investigate the temporal evolution of the non-sphericity and how the non-sphericity depends on the slip length and the equilibrium contact angle. The early time dynamics of the transient ridge is studied in section \ref{tran_ridge}. In section \ref{pertur}, we give explanations for the behavior of the non-sphericity, in terms of the flow structure and the spreading of the ridge.

\subsection{Interfacial profile evolution and non-sphericity of the profiles}\label{profilevol}

Here we briefly consider the droplet geometries studied in \citet{McGraw16}, namely an initial spherical cap with $\theta_i$ = $10\,^\circ$. The equilibrium contact angle $\theta_e= 62\,^\circ$. As discussed in \citet{McGraw16}, the main characteristic feature of the profile evolution is the appearance, or absence, of a transient ridge, defined as the fluid region in between the contact line and the outermost inflection point of the droplet profile (i.e. $\partial^2 \tilde{h}/\partial \tilde{r}^2|_{\tilde{r}=\tilde{r}_{\rm inf}}=0$). The ridge may develop to a global bump, characterized by a maximum in the height profile at $r \neq 0$. The properties of the global bump will be discussed in section \ref{bump}. We first look at two different rescaled slip lengths, $\tilde{b}$ = 0.46 and 23.2, which respectively demonstrate the formation or not of a transient ridge. The evolution of the free interface profiles $\tilde{h}(\tilde{r},\tilde{t})$ are shown in figure \ref{profiles}(a) for $\tilde{b}$ = 0.46, and figure \ref{profiles}(b) for $\tilde{b}$ = 23.2. The main difference between the two cases is that, for $\tilde{b}$ = 0.46, the profile around the center does not change appreciably at early times. The fluid accumulates in a rim as the contact line moves towards the center of the droplet, and forms a transient ridge.  In contrast, for $\tilde{b}$ = 23.2, the height of the interface profile at the center of the droplet increases at early times due to elongational flow \citep{McGraw16}. No ridge is developed in this case.

\begin{figure}
\begin{center}
\includegraphics[width=1.0\textwidth]{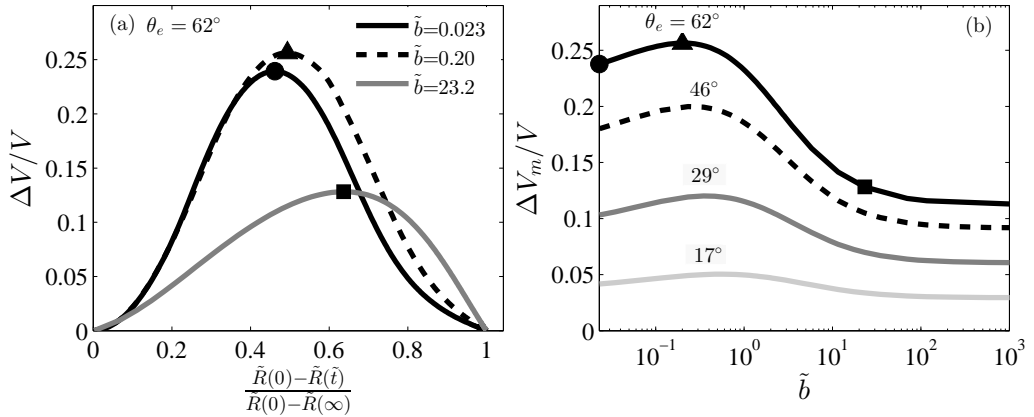}
\caption{(a) The non-sphericity, $\Delta V/V$, versus the rescaled contact line displacement $(\tilde{R}(0)-\tilde{R}(\tilde{t}))/(\tilde{R}(0)-\tilde{R}(\infty))$ for $\theta_e=62\,^\circ$. The maxima are indicated by solid symbols. (b) The maximum of $\Delta V/V$ shown in (a), $\Delta V_m/V$, as a function of slip length $\tilde{b}$ for different equilibrim contact angles $\theta_e$. For both (a) and (b), the intitial contact angle $\theta_i=10\,^{\circ}$.}\label{dVm_slip}
\end{center}
\end{figure}

The transient profiles of the droplets shown in figure \ref{profiles} deviate significantly from the shape of a spherical cap. To quantify the non-sphericity of the droplet, we determine the spherical cap of profile $\tilde{z}=\tilde{S}(\tilde{r},\tilde{t};\tilde{\rho},\tilde{S}_0)$ that best fits the profile of the droplet; $\tilde{S}$ is given implicitly by $\tilde{\rho}^2 = \tilde{r}^2+(\tilde{S}-\tilde{S}_0)^2$, where $\tilde{S}_0$ is the vertical shift of the sphere center while $\tilde{\rho}$ is its radius of curvature. More precisely, for each time $\tilde{t}$ we introduce the observable $\Delta V$ defined by
\begin{equation}
\Delta V = \min\limits_{\tilde{\rho},\tilde{S}_0}\left(\int_0^\infty d\tilde{r}\ 2\pi \tilde{r}|\tilde{h}(\tilde{r},\tilde{t}) - \tilde{S}(\tilde{r},\tilde{t};\tilde{\rho},\tilde{S}_0)|\right)\ , 
\end{equation}
under the constraint of identical total volumes: 
\begin{align}
V=\int_0^\infty d\tilde{r}\ 2\pi \tilde{r}\tilde{h}(\tilde{r},\tilde{t}) = \int_0^\infty d\tilde{r}\ 2\pi \tilde{r}\tilde{S}(\tilde{r},\tilde{t};\tilde{\rho},\tilde{S}_0)\ . 
\end{align}
Note that 
\begin{eqnarray}
\begin{array}{l l}
\tilde{h}(\tilde{r},\tilde{t})=0 & \quad \textrm{for $\tilde{r}>\tilde{R}(\tilde{t})$ ,} \\
\tilde{S}(\tilde{r},\tilde{t};\tilde{\rho},\tilde{S}_0)=0 & \quad \textrm{for $\tilde{r}>\tilde{R}_{\rm cap}(\tilde{t};\tilde{\rho},\tilde{S}_0)$ },
\end{array}
\end{eqnarray}
where $\tilde{R}(\tilde{t})$ and $\tilde{R}_{\rm cap}(\tilde{t};\tilde{\rho},\tilde{S}_0)$ are the contact line radius of the droplet and the spherical cap respectively.
Clearly $\Delta V$  changes throughout the droplet evolution. 

In figure \ref{dVm_slip} (a), $\Delta V$ rescaled by the volume of the droplet is plotted as a function of the contact line displacement $ \tilde{R}(0)-\tilde{R}(\tilde{t}) $ normalized by the total displacement $\tilde{R}(0)-\tilde{R}(\infty)$ for $\tilde{b}=0.023$, $0.20$ and $23.2$. For all three cases, $\Delta V/V$ is zero at $\tilde{t}=0$ and at equilibrium because of the spherical cap shape of the droplets. During the evolution, the non-sphericity attains a maximum. This maximal non-sphericity, $\Delta V_m/V$, occurs at smaller contact line displacements for smaller slip lengths. 

A full investigation of $\Delta V_m/V$ as a function of $\tilde{b}$ is shown in figure \ref{dVm_slip}(b) for various $\theta_e$. We observe that this maximal nonsphericity of the droplet evolution is \emph{non-monotonic} with $\tilde{b}$ for all $\theta_e$ investigated. We note furthermore the presence of a well defined maximum at a slip length that we denote $\tilde{b}_m(\theta_e)$. For $\tilde{b}>\tilde{b}_m$, $\Delta V_m/V$ decreases with $\tilde{b}$ and asymptotically saturates to a finite $\Delta V_m(\infty)/V$. For $\tilde{b}<\tilde{b}_m$, $\Delta V_m/V$ decreases with decreasing $\tilde{b}$. As expected, the non-sphericity becomes smaller as the equilibrium contact angle $\theta_e$ approaches the initial contact angle $\theta_i$.

\begin{figure}
\begin{center}
\includegraphics[width=0.9\textwidth]{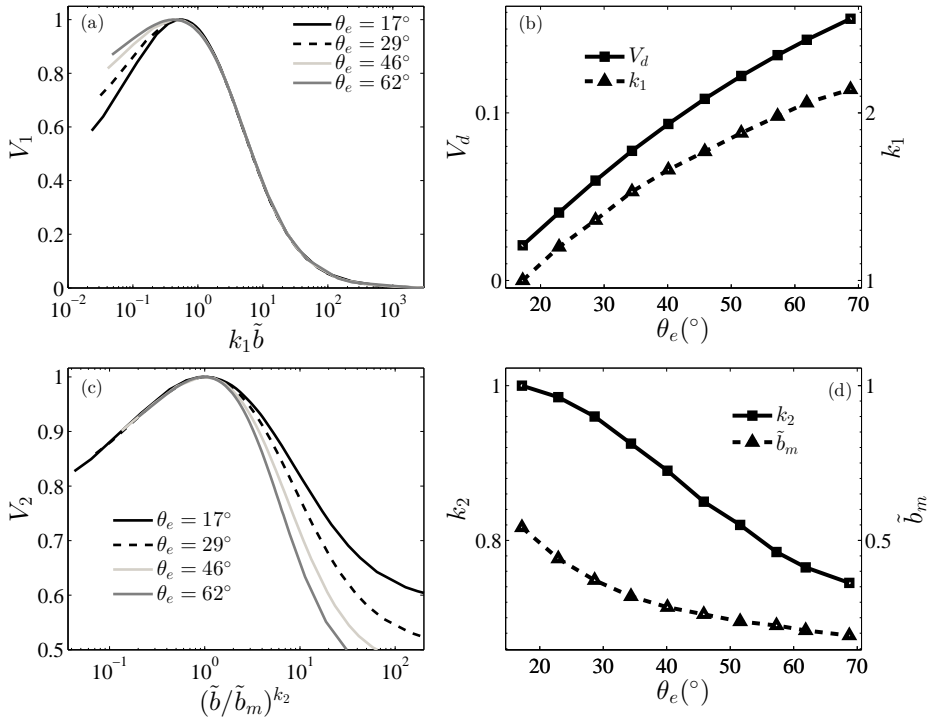}
\caption{After two different rescalings, the data in figure \ref{dVm_slip}(b) collapse into a single function for two different regions, namely $\tilde{b}>\tilde{b}_m$ as shown in (a) and $0.023<\tilde{b}<\tilde{b}_m$ as shown in (c). (a) $V_1$, defined as $(\Delta V_m/V-\Delta V_m(\infty)/V)/V_d$, versus $k_1\tilde{b}$. (b) The dependence of $V_d$ and $k_1$ on the equilibrium contact angle $\theta_e$. (c) $V_2$, defined as $\Delta V_m/V$ rescaled by the $\Delta V_m(\tilde{b}_{m})/V$, versus $(\tilde{b}/\tilde{b}_m)^{k_2}$. (d) $\tilde{b}_m$ and $k_2$ as a function of $\theta_e$. }\label{rescaling}
\end{center}
\end{figure}

The similar features of $\Delta V_m/V$ as a function of $\tilde{b}$ for different equilibrium contact angles $\theta_e$ suggest possible scaling solutions. First, we shift $\Delta V_m/V$ such that all the curves have the same reference level in the full slip limit. Then we rescale the shifted $\Delta V_m/V$ by $V_d \equiv \Delta V_m(\tilde{b}_m)/V-\Delta V_m(\infty)/V$. We hence introduce a rescaled quantity $V_1(\tilde{b})$ as the following:  
\begin{align}
V_1(\tilde{b})&\equiv \frac{1}{V_d}\frac{\Delta V_m(\tilde{b})-\Delta V_m(\infty)}{V}\ . 
\end{align}
The maximum of $V_1$ is unity for any equilibrium contact angles. When plotting $V_1$ versus $\tilde{b}$ multiplied by a scaling factor $k_1(\theta_e)$ in figure \ref{rescaling}(a), we observe that the curves for different $\theta_e$ collapse into a single function for $\tilde{b}>\tilde{b}_m$. The dependence of $V_d$ and $k_1$ on $\theta_e$ is shown in figure \ref{rescaling}(b). Note that $k_1$ is not unique. Multiplying $k_1$ by an arbitrary factor will still collapse all the curves. Here we take $k_1=1$ for $\theta_e=17^{\circ}$.

For $0.023<\tilde{b}<\tilde{b}_m$, a different rescaling is required to reach a collapse of the curves. In figure \ref{rescaling}(c), $V_2$, defined as $\Delta V_m/V$ rescaled by $\Delta V_m(\tilde{b}_m)/V$, is plotted as a function of $(\tilde{b}/\tilde{b}_m)^{k_2}$; a single  curve is thus obtained for $0.023<\tilde{b}<\tilde{b}_m$. This rescaling suggests a relation of the form $V_2/k_2 \sim \textrm{log}(\tilde{b}/\tilde{b}_m)$. Such a logarithmic relation is reminiscent of the weak slip models for nonequilibrium droplets~\citep{degennes85RMP,C86}, in which the contact line dynamics and the interface profile also depend on the slip length logarithmically. The dependence of $\tilde{b}_m$ and $k_2$ on $\theta_e$ is shown in figure \ref{rescaling}(d). The different rescalings for $\tilde{b}<\tilde{b}_m$ and $\tilde{b}>\tilde{b}_m$ indicate the existence of different regimes of the droplet retraction dynamics, which we now describe.

%

\subsection{The transient ridge and early time dynamcis }\label{tran_ridge}

\begin{figure}
\begin{center}
\includegraphics[width=0.9\textwidth]{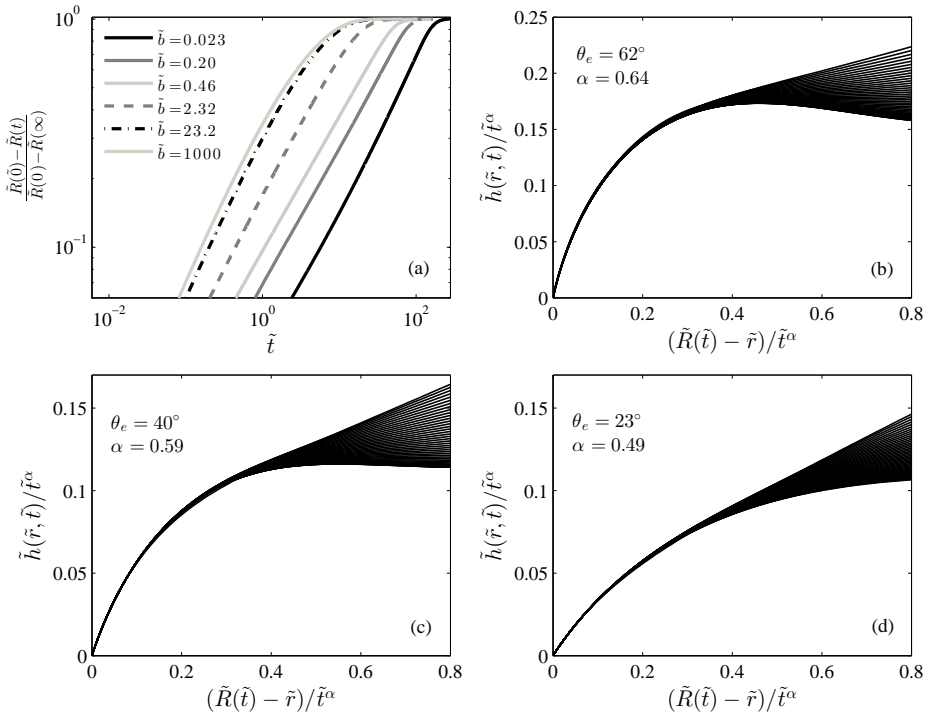}
\caption{(a) Rescaled contact line displacement $(\tilde{R}_0-\tilde{R}(\tilde{t}))/(\tilde{R}_0-\tilde{R}_{\infty})$ as a function of $\tilde{t}$ in log-log scale for different $\tilde{b}$ and fixed $\theta_e=62\,^\circ$. The early time data for intermediate slip lengths around $\tilde{b}_m$ can be described by a power law. A single power law becomes less pronounced for large and small slip lenghts away from $\tilde{b}_m$.  (b), (c) and (d): Rescaled profiles for $\theta_e=62\,^\circ$ and $2< \tilde{t} < 30$ in (b), $\theta_e=40\,^\circ$ and $6<\tilde{t}<58$ in (c) and $\theta_e=23\,^\circ$ and $6<\tilde{t}<116$ in (d). $\tilde{b}=0.46$ in all cases.}\label{earlytimes}
\end{center}
\end{figure}

As one can observe from the profile evolution in figure \ref{profiles}(a), the deviation from a spherical cap of the droplet profiles for intermediate and small slip lengths is related to the formation of the transient ridge. It is thus important to examine the growth of the ridge once the contact line has started to move. We first look at the motion of the contact line. To resolve the contact line motion for early times, we investigate the rescaled contact line displacement $\mathcal{R}(\tilde{t}) \equiv (\tilde{R}(0)-\tilde{R}(\tilde{t}))/(\tilde{R}(0)-\tilde{R}(\infty))$. For given $\theta_i=10\,^\circ$ and $\theta_e=62\,^\circ$, $\mathcal{R}(\tilde{t})$ as a function of time is plotted in figure \ref{earlytimes}(a) in log-log scale for different $\tilde{b}$. For large $\tilde{b}$, the slope of the curves decreases with time. A power law is observed for intermediate slip lengths in the vicinity of the slip length $\tilde{b}_m$ corresponding to the maximal non-sphericity, $\Delta V_m$. We recall that $\tilde{b}_m=0.21$ for $\theta_e=62\,^\circ$. For example, for $\tilde{b}=0.46$ and $0.12<\tilde{t}<30$, the relation, i.e. $\mathcal{R}\sim \tilde{t}^\beta$, describes the data with $\beta=0.59$. The power law relation becomes less pronounced when decreasing $\tilde{b}$ for $\tilde{b}<\tilde{b}_m$. For $\tilde{b}=0.023$, the curve is seen to bend upward with time. Given the power law relation, it is instructive to investigate whether the interface profiles near the contact line can be described by a similarity solution. 

We observe that the local angle of the interface decreases monotonically from $\theta_e$ at the contact line position to around $\theta_i$ at some distance from the contact line in the rim region. Based on this information, we track the coordinates of the point at the free interface with local angle $(\theta_e+\theta_i)/2$. The cylindrical coordinates of this point is denoted as $(\tilde{r}_1, \tilde{h}_1)$. We then investigate how these quantities scale with time. We define $\mathcal{R}_1\equiv(\tilde{R}(\tilde{t})-\tilde{r}_1(\tilde{t}))/(\tilde{R}(\infty)-\tilde{r}_1(\infty))$ and $\mathcal{H}_1\equiv\tilde{h}_1(\tilde{t})/\tilde{h}_1(\infty)$, recalling that $\tilde{R}(\tilde{t})$ is the contact line position as a function of time. We find that both $\mathcal{R}_1$ and $\mathcal{H}_1$ follow power laws in the early time, i.e. $\tilde{t}<30$. The exponents are found to be 0.65 for $\mathcal{R}_1$, and $0.63$ for $\mathcal{H}_1$. Note that these exponents are slightly larger than the exponent, 0.59, obtained for the rescaled contact line position. From the exponents we obtained, it is reasonable to assume a similarity solution of the form $\tilde{h}=\tilde{t}^{\alpha}f((\tilde{R}(\tilde{t})-\tilde{r})/\tilde{t}^{\alpha})$ with $\alpha=0.64$. Indeed all the rescaled interfacial profiles for $2< \tilde{t} < 30$  collapse properly into a single curve for the rim region as shown in figure \ref{earlytimes}(b).  Computing for other equilibrium contact angles and $\tilde{b}=0.46$, the exponents $\alpha$ are found to be 0.59 for $\theta_e=40\,^\circ $ and 0.49  for $\theta_e=23\,^\circ $. The corresponding rescaled profiles are shown in figure \ref{earlytimes}(c) and (d). The exponent 0.49 for the case of $\theta_e=23\,^\circ$, in which the interfacial slope is small, is close to the $\alpha=1/2$ scaling predicted from the lubrication calculation when the dissipation is dominated by the friction at the substrate (\cite{McGraw16}).

\subsection{Spreading of a ridge by shear flow}\label{pertur}

\begin{figure}
\begin{center}
\includegraphics[width=0.9\textwidth]{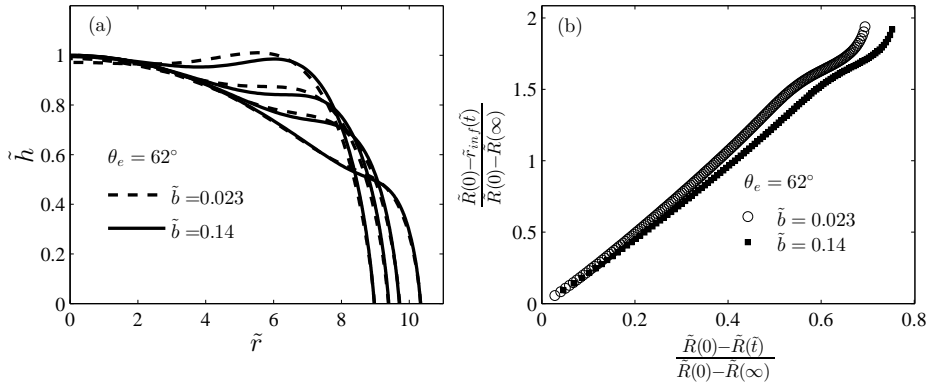}
\caption{(a) Droplet profiles for $\tilde{b}=0.023$ and $0.14$. The profiles are compared  when they have the same contact line position. (b) The rescaled displacement of the inflection point $(\tilde{R}(0)-\tilde{r}_{\rm inf}(\tilde{t}))/(\tilde{R}(0)-\tilde{R}(\infty))$ versus the rescaled contact line displacement $(\tilde{R}(0)-\tilde{R}(\tilde{t}))/(\tilde{R}(0)-\tilde{R}(\infty))$.}\label{spread_wave}
\end{center}
\end{figure}

In this section, we provide explanations for the non-monotonic behavior of the non-sphericity, in terms of the flow structure and the spreading of the ridge.

The vanishing of the transient ridge for large slip lengths has already been discussed in \citet{McGraw16}. For large $\tilde{b}$ (i.e. $\tilde{b}\gg \tilde{b}_m $), low friction at the substrate promotes an elongational flow which affects the whole droplet in a very short time. Therefore, the central height of the droplet increases even at early times due to the upward flow in the center. This prevents mass accumulation at the edge of the droplet. When approaching $\tilde{b}_m$ from large $\tilde{b}$, the elongational flow becomes less dominant. Mass is thus accumulated in the rim while the contact line is moving towards the droplet center. As a consequence a pronounced transient ridge is observed, and the non-sphericity $\Delta V_m/V$, become more strong when decreasing $\tilde{b}$ for $\tilde{b}>\tilde{b}_m$. As shown in section \ref{tran_ridge}, the ridge profiles in the early times can be described by similarity solutions for $\tilde{b}$ close to $\tilde{b}_m$. For small equilibrium contact angles, the similarity solutions can be obtained from the intermediate slip lubrication model in which the dissipation by the friction at the substrate becomes dominant\citep{McGraw16}. That means for $\tilde{b}$ around $\tilde{b}_m$, the  elongational flow and shear flow inside the droplet play relatively minor roles for the dynamics. 

When further decreasing $\tilde{b}$ from $\tilde{b}_m$, the non-sphericity becomes less pronounced. For those small $\tilde{b}$ cases, the flow is more confined to the contact line region and presents a vertical parabolic profile associated with strong shear dissipation. In addition, similarity solutions cannot describe the early ridge profiles anymore. Instead, the question of how much mass is accumulated at the ridge depends on the contact line speed and how fast the mass is redistributed to the central part of the droplet by shear flow. This type of mass redistribution can be observed from the spreading of the ridge. One can imagine a situation when a contact line is pinned from a certain moment, the accumulated mass then has enough time to redistribute to the central part of the droplet and the development of a pronounced global ridge is avoided. Along this line of reasoning, we can understand the decrease of $\Delta V_m/V$ with decreasing $\tilde{b}$. The characteristic speed of the contact line decreases logarithmically with decreasing $\tilde{b}$ for small $\tilde{b}$ \citep{McGraw16}, which means that the disturbance at the contact line will have more time to spread for smaller $\tilde{b}$. This result is demonstrated in figures \ref{spread_wave}(a) and (b) for the case of $\theta_e=62\,^\circ$. In figure \ref{spread_wave}(a), several interface profiles are shown for $\tilde{b}=$0.023 and 0.14. For both cases, the slip lengths are smaller than $\tilde{b}_m$, so the shear dissipation dominates over the elongational one. The profiles are compared for the same contact line position. One clearly sees that the ridge spreads wider for the smaller slip length, namely $\tilde{b}=0.023$.

From the profiles of figure \ref{spread_wave}(a), we observe an outermost inflection point where $d^2\tilde{h}/d\tilde{r}^2=0$. The position of the inflection point $\tilde{r}_{\rm inf}(\tilde{t})$ is used to characterize the extent of the ridge. The displacement of this inflection point $(\tilde{R}(0)-\tilde{r}_{\rm inf}(\tilde{t}))$ normalized by $(\tilde{R}(0)-\tilde{R}(\infty))$ is plotted as a function of the rescaled contact line displacement in figure \ref{spread_wave}(b). It is found that first, the inflection point moves faster than the contact line for both cases, and second, for the same contact line position, the inflection point displaces more for $\tilde{b}=0.023$ compared to $\tilde{b}=0.14$. This result shows again that mass is redistributed over a wider extent for the smaller slip length, $\tilde{b}=0.023$. Hence the non-sphericity  decreases with decreasing $\tilde{b}$. Although we are numerically limited to the smallest $\tilde{b}$ = 0.023, from the trends shown in figure \ref{dVm_slip}(b), we expect that $\Delta V_m/V$ diminishes in the limit of vanishing $\tilde{b}$. Our study thus indicates a crossover from a non-quasistatic regime to a quasistatic regime when $\tilde{b}$ is small.

\subsection{Characteristic of the global bump}\label{bump}
\begin{figure}
\begin{center}
\includegraphics[width=0.9\textwidth]{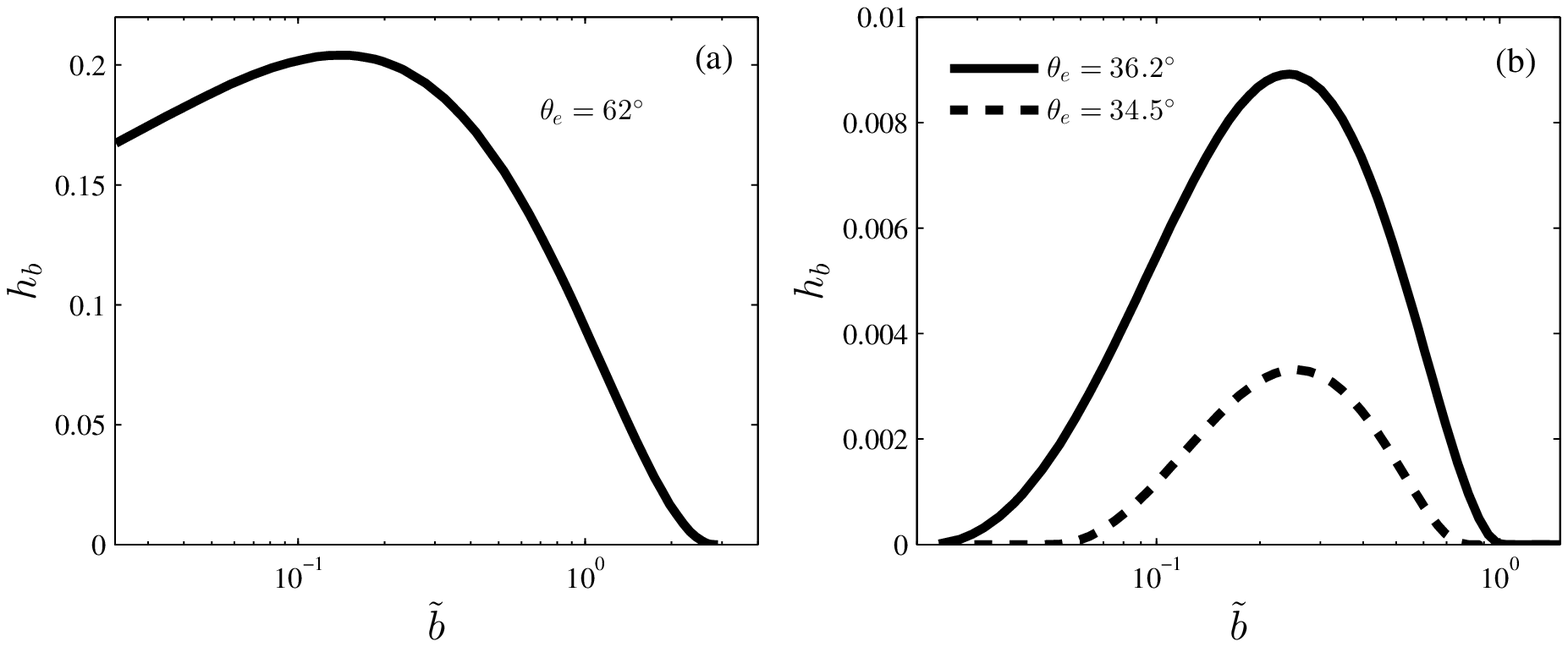}
\caption{The global bump height $h_b$ versus $\tilde{b}$ for  $\theta_i=10\,^\circ$. The equilibrium contact angle $\theta_e=62\,^\circ$ in (a), $\theta_e=34.5\,^\circ$ or $36.2\,^\circ$ in (b).}\label{dhb_third}
\end{center}
\end{figure}

In this section we discuss the properties of the global bump, which reflects the global feature of the droplet profile. Understanding of this feature might be useful for droplet manipulations in micro- and nanofluidics. One can characterize the size of the global bump by measuring the difference between the maximum height of the profile and the central height of the droplet, which we refer to as the global bump height. Like the non-sphericity $\Delta V_m/V$, the global bump height attains a maximum value, denoted as $h_b$ , throughout the profile evolution. A typical behavior of $h_b$ as a function of slip length $\tilde{b}$ is shown in figure \ref{dhb_third}(a) for $\theta=62\,^\circ$. The behaviors of $\Delta V_m/V$ and $h_b$ are similar. The maximum bump height $h_b$ is a non-monotonic function of the slip length and the maximum of $h_b$ occurs at almost the same $\tilde{b}$ as for $\Delta V_m/V$.  For $\tilde{b}$ larger than the point of the maximum, we define the slip length at which $h_b$ goes to zero as $\tilde{b}\equiv \tilde{b}^*$, which equals to 2.81 for the case of $\theta_e=62^{\circ}$. No transient global bump is observed for $\tilde{b}>\tilde{b}^*$.     

\begin{figure}
\begin{center}
\includegraphics[width=0.8\textwidth]{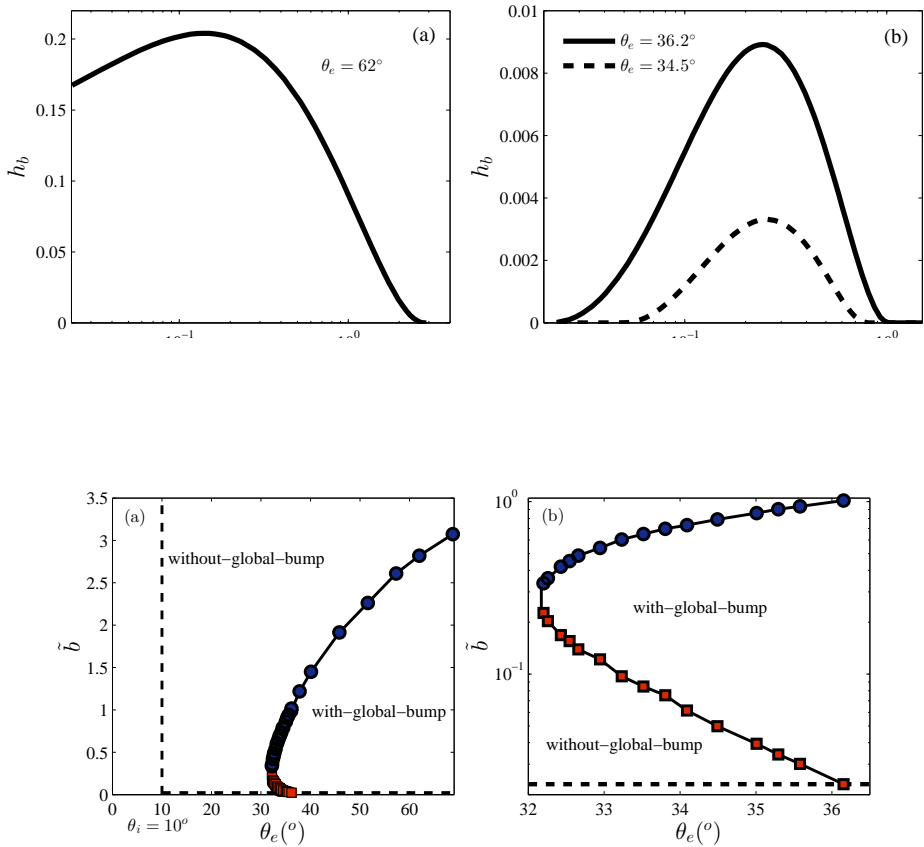}
\caption{(a) Phase diagram showing the with-global-bump region and no-global-bump region. Vertical dotted line indicates the initial contact angle $\theta_i=10\,^\circ$.  Horizontal dotted line indicate the smallest $\tilde{b}$ we have computed for, which is 0.023. Circles are $b^*$ as a function of $\theta_e$. Squares are $\tilde{b}_L^*$ as a function of $\theta_e$. (b) Zoom on the region where $\tilde{b}_L^*$ is computed. Note that $\tilde{b}$ (y-axis) is in log-scale.}\label{all_phase_diag}
\end{center}
\end{figure}

Accessing more values of the equilibrium contact angle, we find another transition. For example, for $\theta_e=34.5^{\circ}$, $\tilde{b}^*=0.79$. When $\tilde{b}$ is further decreased from $\tilde{b}^*,$ we observe a transition from ``with-global-bump'' to ``without-global-bump'' at a certain $\tilde{b}$, which is denoted as $\tilde{b}_L^*$ here, and equals 0.050 in this case. This result means a transient global bump exists when $\tilde{b}_L^* < \tilde{b} < \tilde{b}^*$. This interesting behavior can be observed clearly in figure \ref{dhb_third}(b) where the bump height $h_b$ is plotted as a function of $\tilde{b}$. To summarize the results, a phase diagram is plotted in figure \ref{all_phase_diag} to reveal whether a global transient bump can be observed or not, for the specific case of $\theta_i = 10\,\,^\circ$. When   $\theta_e$ is close to $\theta_i$, namely $\theta_e<32.1^{\circ}$, no global bump appears for any value of $\tilde{b}$. In these cases, a ridge is observed at the early stage for small slip lengths. However, a global bump (with a profile maximum not at $r=0$) does not form because the initial and the final droplet shapes are too similar. In figure \ref{all_phase_diag}(b), one can observe a bifurcation starts from $\theta_e=32.1^{\circ}$. Although the second transition is not observed for $\theta_e >36.2^{\circ}$ due to numerical limitations, we expect the bump to diminish in magnitude also for very small slip lengths in this case; the decrease of $h_b$ in figure \ref{dhb_third}(a) for small $\tilde{b}$ supports this argument. Nevertheless, the slip length below which the global bump disappears is expected to be extremely small if the difference between the initial contact angle and the equilibrium contact angle is large. A recent study has demonstrated that a pronounced global bump exists in the dewetting of very flat droplets ($h_0/R(0)\approx 0.02$) even though the slip length is very small ($\tilde{b}\approx 10^{-5}$) \citep{AMJ2016}. In such cases, the transient global bump itself can be treated as quasistatic, as is the case in dewetting rims of thin liquid films \citep{RBR91,SnE10,Rivetti2015}.

\section{Conclusion}\label{con}
In this article, we study numerically the dewetting of a droplet, with an initial contact angle smaller than the equilibrium contact angle, using the boundary element method for axisymmetric Stokes flow. We impose the Navier-slip boundary condition at the solid/liquid boundary, and a time-independent equilibrium contact angle at the contact line position. The profile evolution is computed for a wide range of slip lengths ($2.3\times10^{-2}<\tilde{b}<10^4$). For all our computations, the transient droplet profiles are found to deviate significantly from a spherical cap. One the other hand, one might expect the droplet to appear as a spherical cap shape throughout the whole evolution when $\tilde{b}$ is small enough under the assumption of quasistatic approach used in the majority of large scale contact line motion problems. To bridge the gap between our computational results and the expectation from the quasistatic approach in the small slip length limit, we investigate the non-sphericity of the dewetting droplet. We find that when decreasing the slip length, the typical non-sphericity first increases, reaches a maximum at a characteristic slip length $\tilde{b}_m$, and then decreases. This non-monotonic behavior is found for all of the equilibrium contact angles investigated in this study, from $17\,^\circ\leq\theta_e\leq 69\,^\circ$. 

The dependence of the non-sphericity on the slip length for different equilibrium contact angles can be described by two universal relations, one for $\tilde{b}>\tilde{b}_m$ and the other one for $0.023<\tilde{b}<\tilde{b}_m$. This result indicates the existence of different flow structures depending on the value of $\tilde{b}$. For $\tilde{b}\gg\tilde{b}_m$, the flow is dominated by the elongational flow \citep{McGraw16}. Around $\tilde{b}_m$, the dissipation is dominated by the friction at the substrate as shown by the similarity solutions for the rim profile evolution at early times. When $\tilde{b}<\tilde{b}_m$, shear flow becomes more important. We explain the decrease of the non-sphericity with decreasing $\tilde{b}$ in terms of the  spreading of the ridge and the contact line velocity. For smaller slip lengths, the accumulated mass due the movement of the contact line is redistributed to a wider extent, thus the droplet profile is closer to a spherical cap.   

Although our numerical computations are limited to the smallest $\tilde{b}=0.023$ we can access, the trend of  the non-sphericity for $\tilde{b}<\tilde{b}_m$ implies that the transient droplet profile will be close to a spherical-cap shape when $\tilde{b}$ is very small, consistent with the expectation from the quasistatic approach. Our study thus brings a first prediction on the connection between the quasistatic and non-quasistatic regimes of droplet dewetting.  
\newline

The authors thank Simon Maurer, Michael Benzaquen, Elie Rapha\"el, and Karin Jacobs for a previous joint study on the topic. They thank the Alexander von Humboldt Foundation, NSERC of Canada and the DFG (Germany) for financial support. The authors also acknowledge financial support from the Global Station for Soft Matter -- a project of Global Institution for Collaborative Research and Education at Hokkaido University. JDM was supported by LabEX ENS-ICFP: ANR-10-LABX-0010/ANR-10-IDEX-0001-02 PSL.

\section{Appendix} 
\subsection{Expressions of $\bar{G}_{\alpha\beta}$ and $\bar{T}_{\alpha\beta\zeta}$}

 For the axisymmetric Stokes flow problem we study in this article, the boundary integral equation is quoted in equation (\ref{bem2}). Here, we provide the standard expressions of the tensors $\bar{G}_{\alpha\beta}$ and $\bar{T}_{\alpha\beta\zeta}$. Note that different symbols are used for these tensors in the book of \cite{Pozbook}. First, we introduce a function $I_{mn}$, which is defined as 
\begin{eqnarray}
I_{mn}\equiv\frac{4k^m}{(4rr_0)^{m/2}}\int_0^{\pi/2}\frac{(2\cos^2w-1)^n}{(1-k^2\cos^2w)^{m/2}}dw.
\end{eqnarray}
 $k$ is given as 
\begin{eqnarray}
k\equiv\left(\frac{4rr_0}{z_d^2+(r+r_0)^2}\right)^{1/2},
\end{eqnarray}
and $z_d\equiv z-z_0$. Here  $(r, z)$ and $(r_0, z_0)$ are the  coordinates of $\boldsymbol{s}$ and $\boldsymbol{s}_0$ respectively.

For $\bar{G}_{\alpha\beta}$,

\begin{equation}
\bar{G}_{zz}=r(I_{10}+z_d^2I_{30}),
\end{equation}
\begin{equation}
\bar{G}_{zr}=rz_d(sI_{30}-r_0I_{31}),
\end{equation}
\begin{equation}
\bar{G}_{rz}=rz_d(rI_{31}-r_0I_{30}),
\end{equation}
\begin{equation}
\bar{G}_{rr}=r[I_{11}+(r^2+r_0^2)I_{31}-rr_0(I_{30}+I_{32})].
\end{equation}

For $\bar{T}_{\alpha\beta\zeta}$,

\begin{eqnarray}
\bar{T}_{zzz}=-6rz_d^3I_{50},
\end{eqnarray}
\begin{eqnarray}
\bar{T}_{zzr}=\bar{T}_{zrz}=-6rz_d^2(rI_{50}-r_0I_{51}),
\end{eqnarray}
\begin{eqnarray}
\bar{T}_{zrr}=-6rz_d(r_0^2I_{52}+r^2I_{50}-2rr_0I_{51}),
\end{eqnarray}
\begin{eqnarray}
\bar{T}_{rzz}=-6rz_d^2(rI_{51}-r_0I_{50}),
\end{eqnarray}
\begin{eqnarray}
\bar{T}_{rzr}=\bar{T}_{rrz}=-6rz_d[(r^2+r_0^2)I_{51}-rr_0(I_{50}+I_{52})],
\end{eqnarray}
\begin{eqnarray}
\bar{T}_{rrr}=-6r[r^3I_{51}-r^2r_0(I_{50}+2I_{52})+rr_0^2(I_{53}+2I_{51})-r_0^3I_{52}].
\end{eqnarray}

The tensors $\bar{G}_{\alpha\beta}$ and $\bar{T}_{\alpha\beta\zeta}$ have singular points at $\boldsymbol{s} = \boldsymbol{s}_0$ and $\boldsymbol{s}=0$. Around these points, the boundary integral equation (\ref{bem2}) is performed analytically by expanding $\bar{G}_{\alpha\beta}$ and $\bar{T}_{\alpha\beta\zeta}$ in series \citep{vanLangerich12JFM}.

\bibliographystyle{jfm}
\bibliography{new_all_ref}

\end{document}